\documentclass[aps,pra,reprint,superscriptaddress,floatfix,footinbib,longbibliography]{revtex4-1}

\usepackage{dcolumn}
\usepackage{graphicx}
\usepackage{amsmath,amssymb}
\usepackage{bm}
\usepackage{epstopdf}
\usepackage{epsfig}

\usepackage[T1]{fontenc}
\usepackage[applemac]{inputenc}
\usepackage{lmodern}
\usepackage[english]{babel}

\usepackage{ae}
\usepackage{tikz}
\usetikzlibrary{arrows}

\usepackage{color}
\usepackage{url}

\usepackage[colorlinks]{hyperref}
\hypersetup{%
	plainpages=true,
	breaklinks=true,
	hypertexnames=false,
	pageanchor=true,
	colorlinks=true,
	linkcolor={blue},
	citecolor={magenta},
	urlcolor={blue},
	anchorcolor={black}
}

\usepackage{mleftright} %to get rid of the extra space around parentheses with \left and \right: write \mleft, \mright instead

\newcommand{\figref}[1]{\mbox{Fig.~\ref{#1}}}

\newcommand{\secref}[1]{\mbox{Sec.~\ref{#1}}}

\renewcommand{\eqref}[1]{\mbox{Eq.~(\ref{#1})}}

\newcommand{\ket}[1]{|#1\rangle}
\newcommand{\braket}[2]{\langle #1|#2\rangle}
\newcommand{\ketbra}[2]{\mleft| #1 \rangle \langle #2 \mright|}
\newcommand{\brakket}[3]{\langle #1 | #2 | #3 \rangle}
\newcommand{\expec}[1]{\mleft\langle #1 \mright\rangle}

\newcommand{\comm}[2]{\mleft[ #1, #2 \mright]}

\newcommand{\sz}{\hat \sigma_z}
\newcommand{\sx}{\hat \sigma_x}
\newcommand{\sm}{\hat \sigma_-}
\renewcommand{\sp}{\hat \sigma_+}

\newcommand{\abssq}[1]{\mleft| #1 \mright|^2}

\newcommand{\rd}{\ensuremath{\mathrm{d}}} %For derivatives and integrals

\newcommand{\nn}{\nonumber}

\newcommand{\be}{\begin{equation}}
	\newcommand{\ee}{\end{equation}}
\newcommand{\bea}{\begin{eqnarray}}
	\newcommand{\eea}{\end{eqnarray}}

\begin{document}

\title{Photodetection probability in quantum systems with \\ arbitrarily strong light-matter interaction}
%\title{Photodetection probability in systems with arbitrarily strong light-matter interaction}
%\title{Photodetection probability in Interacting Light-Matter Systems}

\author{Omar Di Stefano}
%\affiliation{MIFT - Dipartimento di Scienze Matematiche e Informatiche Scienze Fisiche e Scienze della Terra, Universit\`{a} di Messina, I-98166 Messina, Italy}
\affiliation{Center for Emergent Matter Science, RIKEN, Saitama 351-0198, Japan}

%\author{Roberto Stassi}
%%\affiliation{MIFT - Dipartimento di Scienze Matematiche e Informatiche Scienze Fisiche e Scienze della Terra, Universit\`{a} di Messina, I-98166 Messina, Italy}
%\affiliation{Center for Emergent Matter Science, RIKEN, Saitama 351-0198, Japan}
%
%\author{Luigi Garziano}
%\affiliation{University of Southampton, Southampton, SO17 1BJ, United Kingdom }
%
%\author{Vincenzo Macr\'{i}}
%\affiliation{MIFT - Dipartimento di Scienze Matematiche e Informatiche Scienze Fisiche e Scienze della Terra, Universit\`{a} di Messina, I-98166 Messina, Italy}
%%\affiliation{Center for Emergent Matter Science, RIKEN, Saitama 351-0198, Japan}
%

\author{Anton Frisk Kockum}
\affiliation{Center for Emergent Matter Science, RIKEN, Saitama 351-0198, Japan}

\author{Alessandro Ridolfo}
%\affiliation{MIFT - Dipartimento di Scienze Matematiche e Informatiche Scienze Fisiche e Scienze della Terra, Universit\`{a} di Messina, I-98166 Messina, Italy}
\affiliation{Center for Emergent Matter Science, RIKEN, Saitama 351-0198, Japan}

\author{Salvatore Savasta}
\affiliation{MIFT - Dipartimento di Scienze Matematiche e Informatiche Scienze Fisiche e Scienze della Terra, Universit\`{a} di Messina, I-98166 Messina, Italy}
\affiliation{Center for Emergent Matter Science, RIKEN, Saitama 351-0198, Japan}

\author{Franco Nori}
\affiliation{Center for Emergent Matter Science, RIKEN, Saitama 351-0198, Japan}
\affiliation{Physics Department, The University of Michigan, Ann Arbor, Michigan 48109-1040, USA}

\date{\today}

\begin{abstract}

Cavity-QED systems have recently reached a regime where the light-matter interaction strength amounts to a non-negligible fraction of the resonance frequencies of the bare subsystems. In this regime, it is known that the usual normal-order correlation functions for the cavity-photon operators fail to describe both the rate and the statistics of emitted photons. Following Glauber's original approach, we derive a simple and general quantum theory of photodetection, valid for arbitrary light-matter interaction strengths. Our derivation uses Fermi's golden rule, together with an expansion of system operators in the eigenbasis of the interacting light-matter system, to arrive at the correct photodetection probabilities. We consider both narrow- and wide-band photodetectors. Our description is also valid for point-like detectors placed inside the optical cavity. As an application, we propose a gedanken experiment confirming the virtual nature of the bare excitations that enrich the ground state of the quantum Rabi model.

\end{abstract}

\maketitle

\section{Introduction}

The problem of the theoretical description of the photon-detection process was addressed by Glauber in Ref.~\cite{Glauber1963}. In this pioneering work, he formulated the quantum theory of photodetection and optical coherence. This theory is central to all of quantum optics and has occupied a key role in understanding light-matter interactions. In order to discuss measurements of the intensity of light, Glauber described the photon detector as a system that functions by absorbing quanta and registering each such absorption process, e.g., by the detection of an emitted photoelectron. In particular, Glauber defined an ideal photon detector as ``a system of negligible size (e.g., of atomic or subatomic dimensions) which has a frequency-independent photoabsorption probability''. Since the photoabsorption is independent of frequency, such an ideal, small detector, situated at the point $\mathbf{r}$, can be regarded as probing the field at a well defined time $t$. Glauber showed that the rate at which the detector records photons is proportional to $\brakket{i}{\hat E^- (\mathbf{r}, t) \hat E^+ (\mathbf{r}, t)}{i}$, where $\ket{i}$ describes the initial state of the electromagnetic field. The operators $\hat E^\pm (\mathbf{r}, t)$ are the positive- and negative-frequency components of the electromagnetic field operator $\hat E (\mathbf{r}, t) = \hat E^+ (\mathbf{r}, t) + \hat E^- (\mathbf{r}, t)$, i.e., the components with terms varying as $e^{- i \omega t}$ for all $\omega > 0$ (positive-frequency components) or as $e^{i \omega t}$ for all $\omega >0$ (negative-frequency components).

In cavity quantum electrodynamics (cavity QED)~\cite{Haroche2006, Haroche2013}, where atoms interact with discrete electromagnetic field modes confined in a cavity, it is often the photons leaking out from the cavity that are detected in experiments. To describe the dynamics of the atoms and the photons in the cavity, it is common to adopt a master-equation approach (see, e.g., Refs.~\cite{Breuer2002, Gardiner2004}). In this approach, the field modes outside the cavity are treated as a heat bath, whose degrees of freedom are traced out when deriving the master equation. As a consequence, the master equation cannot be directly applied to derive the output field that is to be detected. 

This gap between the quantum system and the external detector is typically bridged by input-output theory~\cite{Gardiner1985, Walls2008}, which can be used to determine the effect of the intra-cavity dynamics on the quantum statistics of the output field in a very clear and simple way. In particular, if we limit the discussion to a single cavity mode, with annihilation operator $\hat a$, interacting with an external field and apply the rotating-wave approximation (RWA), it is possible to obtain the output field operator $\hat a_{\rm out} (t)$ as a function of the intra-cavity field $\hat a (t)$ and the input field $\hat a_{\rm in} (t)$ operators~\cite{Gardiner1985, DiStefano2001, Walls2008}:
\be
\hat a_{\rm out} (t) = \hat a_{\rm in} (t) + \sqrt{\kappa} \hat a (t),
\ee
where $\kappa$ is an input-output coupling coefficient describing the cavity loss rate.

In recent years, cavity QED has thrived thanks to an increase in the ability to control light-matter interaction at the quantum level. In particular, owing to the the advances in the detection, generation and emission of photons~\cite{Hadfield2009, Buller2010, Eisaman2011, Sathyamoorthy2016a, Gu2017}, quantum systems are increasingly addressed at the single-photon level. As a consequence, there is a pressing need for a critical analysis of the applicability of the theory of photodetection~\cite{Vogel2006, Ridolfo2012, DelValle2012}. Moreover, photon correlations are now routinely measured in the laboratory and many experiments, ranging from studying effects of strong and ultrastrong light-matter coupling to performing quantum state tomography or monitoring single-photon emitters (see, e.g.,~\cite{Aspect1980, Schrama1991, Press2007, Hennessy2007, Hofheinz2009, Bozyigit2011, Eichler2011, You2011, Ulhaq2012, Stassi2016, Gu2017}), have shown their power in characterizing quantum systems. In addition, photodetection is also used for quantum-state engineering~\cite{Bimbard2010} and quantum information protocols~\cite{Knill2001, Hadfield2009}.

For these complex systems, i.e., realistic atom-cavity systems, the theory of photodetection must be applied with great care because the light-matter interaction may modify the properties of the bare excitations in the system. If the physical excitations in such systems are superpositions of light and matter excitations, it is not immediately clear what really is measured in a photodetection experiment.

More specifically, we observe that the interaction Hamiltonian of a realistic atom-cavity system contains so-called counter-rotating terms, which allow simultaneous creation or annihilation of excitations in both atom and cavity modes (see, e.g., Ref.~\cite{DiStefano2017}). These terms can be safely neglected through the RWA for small atom-cavity coupling rates $g$. However, when $g$ becomes comparable to the resonance frequencies of the atoms or the cavity, the counter-rotating terms manifest, giving rise to a host of interesting effects~\cite{DeLiberato2007, Ashhab2010, Cao2010, Casanova2010, Beaudoin2011, Ridolfo2012, Stassi2013, DeLiberato2014, Sanchez-Burillo2014, Garziano2015, Lolli2015, Cirio2016, Garziano2016, DiStefano2017, Gu2017, Kockum2017, Kockum2017a, Peng2017, Stassi2017}. This ultrastrong coupling (USC) regime is difficult to reach in optical cavity QED, but was recently realized in a variety of solid-state quantum systems~\cite{Gunter2009, Niemczyk2010, Schwartz2011, Scalari2012, Geiser2012, Gambino2014, George2016a, Forn-Diaz2017, Chen2017, Yoshihara2017, Gu2017}. The USC regime is challenging from a theoretical point of view because the total number of excitations in the cavity-emitter system is not conserved (only the parity of the number of excitations is)~\cite{Casanova2010, Braak2011}. 

In the USC regime, it has been shown that the quantum-optical master equation fails to provide the correct description of the system's interaction with reservoirs~\cite{Beaudoin2011}. It was also found~\cite{Ridolfo2012} that a naive application of the standard descriptions of photodetection and dissipation fail for thermal emission from a cavity-QED in the USC regime. In addition, quantum-optical normal-order correlation functions fail to describe photodetection experiments for such systems~\cite{DeLiberato2007}. To understand why an incautious application of Glauber's idea of photodetection together with standard input-output theory will give incorrect results, consider a USC system in its ground state $\ket{G}$. Due to the influence of the counter-rotating terms in the Hamiltonian, $\brakket{G}{\hat a^\dag\hat a}{G} \neq 0$, and since standard input-output theory predicts that $\expec{\hat a^\dag_{\rm out}\hat a_{\rm out}} \propto \expec{\hat a^\dag\hat a}$, this would imply that photons could be emitted from the ground state and then detected, which is unphysical. However, with a proper generalization of input-output theory~\cite{Ciuti2006}, Glauber's idea of photodetection can still be applied to the output from a USC system~\cite{Ridolfo2012}.

In this article, we present a general and simple quantum theory of the photodetection for quantum systems with \textit{arbitrarily strong light-matter interaction}. We show how Glauber's original results for the quantum theory of photodetection can be applied to systems in the USC regime. In contrast to previous works, our approach does \textit{not} require the use of input-output theory and therefore applies also to more general physical situations, where it is \textit{not} possible to measure and/or identify the output photons.

In order to calculate the detection probability of the photoabsorber, we use the more general Fermi's golden rule. As a consequence, our approach can be applied to measurements of field correlations inside an optical resonator. In such a case, it is not possible to use the input-output approach because the interaction strongly modifies the positive- and negative-frequency field components. Their explicit expressions, in fact, contain combinations of the bare creation and destruction photon operators, which cannot be treated separately as would be required in input-output theory. In addition, using the correct positive- and negative-frequency parts of the field dressed by the interaction, we are able to calculate the photodetection probabilities for both narrow- and wide-band photodetectors. 

We observe that a key theoretical issue for systems in the USC regime is the distinction between bare (unobservable) excitations and physical particles that can be detected~\cite{DiStefano2017}. Several works~\cite{Ciuti2006, Ridolfo2012, Stassi2013} have shown that the photons in the ground state are not observable, in the sense that they do not give rise to any output photons that can be observed by standard photon detection. However, other works have shown that the photons in the ground state may be indirectly detected (without being absorbed)~\cite{Garziano2014, Lolli2015, Cirio2017, Endo2017}. The formalism we develop here allow us to investigate the issue of these ground-state photons more deeply than before, elucidating their virtual nature. The conclusions we can draw here also apply to excited states in the USC system, which may contain contributions from virtual photonic and atomic excitations.

This article is organized as follows. In \secref{sec:EP}, we derive the photodetection probability for a photoabsorber coupled to a quantum system, which may have arbitrarily strong light-matter interaction. We then show how to apply this formalism to two representative systems in \secref{sec:Applications}. In \secref{sec:Rabi}, we use the results from \secref{sec:EP} to analyze the nature of photonic and atomic excitations dressing the ground and excited states in a USC system. We conclude in \secref{sec:Conclusions}.

%%%%%%%%%%%%%%%%%%%%%%%%%%%%%%%%%%%%%%%%%%%%%%%%%%%%%%%%%%%%

\section{Excitation probability for a photon detector}
\label{sec:EP}

We consider a generic quantum system with light-matter coupling. This quantum system is weakly coupled to a photo-absorber, which is modelled as a quantum system with a collection of modes at zero temperature. The Hamiltonian $\hat{\mathcal{V}}$ describing the coupling between the light-matter system and the photo-absorber is (we set $\hbar = 1$ throughout this article unless otherwise specified)
\be
\hat{\mathcal{V}} = \sum_n g_n \mleft( \hat c^\dag_n + \hat c_n \mright) \hat O,
\label{eq:HintSystemAbsorber} 
\ee
where $\hat O$ is some operator of the light-matter system, $\hat c_n$ ($\hat c^\dag_n$) is an annihilation (creation) operator for mode $n$ of the photo-absorber, and $g_n$ is the strength with which this mode couples to the light-matter system. Typically, the operator $\hat O$ would be the operator of the intra-cavity electromagnetic field in a cavity-QED setup. However, this is not the only possibility. We could also have a situation where $\hat O$ is an operator belonging to the matter part of the system. As for the operators $\hat c_n$ and $\hat c^\dag_n$, their form will depend on the model of the photo-absorber. If the photo-absorber is a collection of harmonic oscillators, $\hat c_n$ and $\hat c^\dag_n$ are bosonic operators. If the photo-absorber is a generic multilevel quantum system, $\hat c_n = \ketbra{n}{0}$, where $\ket{0}$ is the ground state and $\ket{n}$ the $n$th excited state.
 
The aim of this section is to calculate the excitation probability of the photo-absorber, which initially is in its ground state $\ket{0}$. The matrix element governing this excitation process is $\brakket{F_\alpha}{\hat{\mathcal{V}}}{I}$, where $\ket{I} = \ket{E_i, 0}$ and $\ket{F_\alpha} = \ket{E_k, n}$ are the initial and final states, respectively, of the total system (generic light-matter system plus photo-absorber). Here, we denote the eigenstates of the Hamiltonian $\hat H_{\rm s}$ of the light-matter system by $\ket{E_k}$ ($k = 0, 1, 2, \dots$), and the corresponding eigenvalues by $E_k$, choosing the labelling of the states such that $E_k > E_j$ for $k > j$.
 
Using Fermi's golden rule, summing over the possible final states, the resulting excitation probability per unit of time for the photo-absorber can be expressed as
\bea
\frac{\rd W_i}{\rd t} &=& 2 \pi \sum_{n, k} \abssq{\brakket{E_k, n}{\hat{\mathcal{V}}}{E_i, 0}} \delta \mleft( \omega_n + E_k - E_i \mright)\nn\\ 
&=& 2 \pi \sum_{n, k} g_n^2 \abssq{\brakket{E_k}{\hat O}{E_i}} \delta \mleft( \omega_n + E_k - E_i \mright).
\label{ciccio}
\eea
If the photo-absorber has a continuous spectrum, the sum over absorber modes can be replaced by an integral over the corresponding frequencies:
\bea
\frac{\rd W_i}{\rd t} &=& 2 \pi \int_0^\infty \rd \omega g^2 (\omega) \rho (\omega) \nn\\
&& \times \sum_{k < i} \abssq{\brakket{E_k}{\hat O}{E_i}} \delta \mleft( \omega + E_k - E_i \mright),
\label{eq:dWdtIntegralForm}
\eea
where $\rho(\omega)$ is the density of states of the absorber. Note that we also limited the summation to $k < i$, which follows from the delta-function terms since the continuous spectrum of the absorber only contains positive frequencies $\omega$. Using some further manipulation based on this fact, \eqref{eq:dWdtIntegralForm} can be expressed as
\be
\frac{\rd W_i}{dt} = \int_0^\infty \rd \omega \sum_{k < i} \chi (\omega_{i, k}) \abssq{\brakket{E_k}{\hat O}{E_i}} \delta \mleft( \omega + E_k - E_i \mright),
\label{eq:dWdtIntegralForm_Chi}
\ee
where $\omega_{i, k} = \mleft( E_i - E_k \mright) / \hbar$ and we defined $\chi (\omega) = 2 \pi g^2 (\omega) \rho (\omega)$. To further simplify \eqref{eq:dWdtIntegralForm_Chi}, we define the positive-frequency operator
\be
\hat x^+ = \sum_j \sum_{k < j} x_{kj} \ketbra{E_k}{E_j},
\label{Xp}
\ee
where
\be
x_{kj} = \sqrt{\chi (\omega_{j, k})} O_{kj} = \sqrt{\chi (\omega_{j, k})} \brakket{E_k}{\hat O}{E_j}.
\ee
Since in \eqref{eq:dWdtIntegralForm_Chi} $k < i$, we obtain
\be
\sqrt{\chi (\omega_{i, k})} \brakket{E_k}{\hat O}{E_i} = \brakket{E_k}{\hat x^+}{E_i}.
\label{equiv}
\ee
Inserting \eqref{equiv} into \eqref{eq:dWdtIntegralForm_Chi}, we obtain
\bea
\frac{\rd W_i}{\rd t} &=& \sum_{k} \int_0^\infty \rd \omega \brakket{E_i}{\hat x^-}{E_k} \brakket{E_k}{\hat x^+}{E_i} \delta \mleft( \omega + E_k - E_i \mright) \nn\\
&=& \brakket{E_i}{\hat x^- \hat x^+}{E_i},
\label{P4}
\eea
where we used the identity relation $\sum_k \ketbra{E_k}{E_k} = \hat 1$. 
 
The detector excitation rate is thus proportional to the initial-state expectation value of the Hermitian operator $\hat x^- \hat x^+$. We can extend this result to a more general situation where the initial state is mixed, described by the density matrix $\hat \rho = \sum_j P_j \ketbra{j}{j}$, where $P_j$ is the probability that the initial state is $\ket{j}$. In this case, the excitation probability rate becomes
\be
\frac{\rd W_i}{\rd t} = \expec{\hat x^- \hat x^+},
\ee
where
\be
\expec{\hat x^- \hat x^+} = \text{Tr} \mleft[ \hat \rho \hat x^- \hat x^+ \mright].
\ee
If the frequency dependence of $\chi (\omega)$ can be neglected, we can set $\chi (\omega) \equiv \chi$ and write the excitation probability rate as
\be
\frac{\rd W_i}{\rd t} = \chi \expec{\hat O^- \hat O^+},
\label{P5}
\ee
where $\hat O^+ = \sum_i \sum_{j < i} O_{ji} \ketbra{E_j}{E_i}$.

We now consider the case of a narrow-band photodetector, which only absorbs excitations in a narrow band around a frequency $\omega_{\rm d}$. Setting $\omega_n = \omega_{\rm d}$ and $g_n = g (\omega_{\rm d}) \equiv g$ in \eqref{ciccio}, we obtain
\bea
\frac{\rd W_i}{\rd t} &=& 2 \pi \sum_k g^2 \abssq{\brakket{E_k}{\hat O^+(t)}{E_i}} \delta \mleft( \omega_{\rm d} + E_k - E_i \mright) \nn\\
&=& \int \rd \tau \sum_k g^2 O^-_{ik}(t) O^+_{ki}(t) e^{i \mleft( \omega_{\rm d} + E_k - E_i \mright) \tau}, \quad
\label{ciccio2}
\eea
where
\bea
O^{\pm}_{ki}(t) &=& \brakket{E_k}{\hat O^\pm(t)}{E_i}= \brakket{E_k(t)}{\hat O^\pm}{E_i(t)} \nn\\
&=& e^{-i (E_i-E_k) t}\brakket{E_k}{\hat O^\pm}{E_i}.
\eea
Observing that $O^-_{ki}= O^{+\dag}_{ik}$ and $O^{+\dag}_{ki} (t + \tau) = O^{+ \dag}_{ki} (t) e^{-i (E_i - E_k) \tau}$, \eqref{ciccio2} becomes
\bea
\frac{\rd W_i}{\rd t} &=& \int \rd \tau \sum_k g^2 O^-_{ik} (t) O^+_{ki} (t) e^{i \mleft( \omega_{\rm d} + E_k - E_i \mright)\tau} \nn\\
&=& g^2 \int \rd \tau \sum_k O^-_{ik} (t) O^+_{ki} (t + \tau) e^{i \omega_{\rm d} \tau}.
\label{ciccio3}
\eea
Performing the summation over the possible final states $k$, we finally obtain
\be
\frac{\rd W_i}{\rd t} = g^2 \int \rd \tau e^{i \omega_{\rm d} \tau} \expec{\hat O^- (t) \hat O^+ (t + \tau)}_i,
\label{ciccio4}
\ee
where $\expec{\hat O^-(t) \hat O^+(t+\tau)}_i$ is the two-time expectation value with respect to the initial state.

%%%%%%%%%%%%%%%%%%%%%%%%%%%%%%%%%%%%%%%%%%%%%%%%%%%%%%%%%%%%

\section{Applications}
\label{sec:Applications}

We now show how this formalism for photon detection can be applied to two typical quantum systems with light-matter interaction: cavity QED with natural atoms and superconducting circuits with artifical atoms and microwave photons. Once the interaction Hamiltonian with the correct system and photoabsorber operators have been identified, applying the results from \secref{sec:EP} is straightforward.

%%%%%%%%%%%%%%%%%%%%%%%%%%%%%%%%%%%%%%%%%%%%%%%%%%%%%%%%%%%%

\subsection{An atom as a detector for the electromagnetic field in a cavity}

We first consider the electromagnetic field in a cavity, interacting with arbitrary strength with some quantum system, e.g., one or more natural atoms situated in the cavity. As our photoabsorber, we take an atom that is weakly coupled to the field (and not coupled at all to the quantum system that the cavity interacts with). The interaction Hamiltonian describing the field and the absorber can then be written as
\be
\hat{\mathcal{V}} = - \frac{e}{m} \hat{\mathbf{p}} \cdot \hat{\mathbf{A}},
\ee
where $\hat{\mathbf{p}}$ is the atomic momentum operator, $\hat{\mathbf{A}}$ is the vector potential of the electromagnetic field, $e$ is the charge of the electron orbiting the atom, and $m$ is the mass of the electron. We are adopting the Coulomb gauge and we neglected the $\hat A^2$ term, which is a good approximation in the weak-interaction regime. Using \eqref{ciccio}, labelling the atomic eigenstates by $\ket{n}$ and the energies of these states by $\omega_n$ (we set the energy of the ground state $\ket{0}$ to zero), we obtain an expression for the atomic excitation rate:
\be
\frac{\rd W_i}{\rd t} = \sum_{n, k} \abssq{\brakket{E_k, n}{\hat{\mathcal{V}}}{E_i, 0}} \delta \mleft( \omega_n + E_k - E_i \mright).
\label{Pa1}
\ee

By using the relationship $\comm{\hat{\mathbf{r}}}{\hat H_{\rm d}} = i \hat{\mathbf{r}} / m$, where $\hat H_{\rm d}$ is the Hamiltonian of the photodetector in the absence of the interaction with light, we obtain
\be
\brakket{E_k, n}{\hat{\mathcal{V}}}{E_i, 0} = - i \omega_n \mathbf{d}_n \cdot \brakket{E_k}{\hat{\mathbf{A}}}{E_i},
\label{me}
\ee
where $\mathbf{d}_n = \brakket{n}{\hat{\mathbf{r}}}{0}$. Introducing the matrix element from \eqref{me} into \eqref{Pa1} leads to
\be 
\frac{\rd W_i}{\rd t} = \sum_{n, k} \abssq{-i \omega_n \mathbf{d}_n \cdot \brakket{E_k}{\hat{\mathcal{\mathbf{A}}}}{E_i}} \delta \mleft( \omega_n + E_k - E_i \mright),
\ee
which, after introducing the positive-frequency electric-field operator
\be
\hat{\mathbf{E}}^+ = \sum_m \sum_{j < m} - i \mleft( E_j - E_m \mright) \hat{\mathbf{A}}_{jm} \ketbra{E_j}{E_m},
\label{Ep}
\ee
using the Dirac delta function, and assuming a constant dipole moment $\mathbf{d}_n = \mathbf{d}$ (wide-band detector), can be expressed as
\be  
\frac{\rd W_i}{\rd t} = d_\alpha d_\beta \expec{\hat E_\alpha^- \hat E_\beta^+},
\label{eq:ExcitationRateFinalCavityQED}
\ee
where the Greek letters indicate the cartesian components of the dipole moment and of the electric-field operator, and repeated indices are summed over.

We note again that if the strength of the coupling between the cavity field and the quantum system it interacts with (not the photoabsorber) is arbitrarily large, the positive- and negative-frequency electric-field operators appearing in the final expression for the photodetection probability in \eqref{eq:ExcitationRateFinalCavityQED} may not correspond to the bare creation and annihilation operators $a$ and $a^\dag$ of that field. Instead, the photodetection probability is set by transitions between the eigenstates of the full system (cavity field plus the quantum system it interacts with).

%%%%%%%%%%%%%%%%%%%%%%%%%%%%%%%%%%%%%%%%%%%%%%%%%%%%%%%%%%%

\subsection{Circuit QED}

As our second example, we consider a circuit-QED setup. In circuit QED, artificial atoms formed by superconducting electrical circuits incorporating Josephson junctions can be strongly coupled to $LC$ and transmission-line resonators~\cite{Wallraff2004, You2011, Gu2017}. These circuits can be designed to explore new regimes of quantum optics. In particular, recent circuit-QED experiments~\cite{Yoshihara2017, Forn-Diaz2017} hold the current record for strongest light-matter interaction, having reached not only the USC regime but also the regime of deep strong coupling, where the coupling strength exceeds the resonance frequencies of both the (artificial) atom and the electromagnetic mode(s). Circuit-QED systems are also being used to investigate virtual and real photons in other settings than ultrastrong light-matter interaction~\cite{Nation2012}, e.g., in the dynamical Casimir effect~\cite{Johansson2009, Johansson2010, Wilson2011, Johansson2013}.

\begin{figure}
\centering
\includegraphics[width = \linewidth]{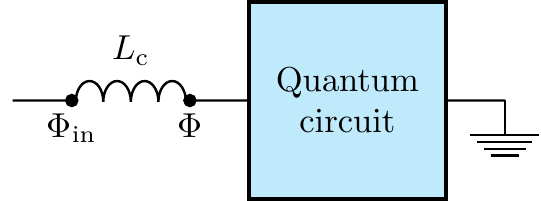}
\caption{Sketch of a circuit-QED system inductively coupled to a photon detector.}
\label{quedexamplex}
\end{figure}

As sketched in \figref{quedexamplex}, we treat our (possibly quite complex) quantum circuit as a ``black box''~\cite{Nigg2012}. The quantum circuit will contain both electromagnetic modes and artificial atom(s), but for our purposes it is sufficient to know how this system as a whole couples to an absorbing photon detector. We assume that the coupling is through an inductor $L_{\rm c}$ that connects a node flux $\Phi$ of the circuit to a node flux $\Phi_{\rm in}$ of the photoabsorber.

From standard circuit quantization methods~\cite{Vool2017}, it follows that the interaction Hamiltonian for our setup is~\cite{Girvin2014}
\be
\hat H_{\rm int} = \frac{1}{2 L_{\rm c}} \mleft( \hat \Phi - \hat \Phi_{\rm in} \mright)^2,
\label{cqed1}
\ee
where the node fluxes have been promoted to quantum operators and thus acquired hats. The operator $\hat \Phi_{\rm in}$ represents the measurement system that we hook up to our quantum circuit; it can be rewritten as a weighted contribution of absorber modes:
\be
\hat \Phi_{\rm in} = \sum_n k_n \mleft( \hat c_n + \hat c^\dag_n \mright).
\label{cqed2}
\ee
Similarly, the flux operator $\hat \Phi$ can be expressed as~\cite{Girvin2014}
\be
\hat \Phi = \sum_m \Phi_{\rm ZPF}^{(m)} \mleft( \hat a_m + \hat a_m^\dag \mright),
\label{cqed3}
\ee
where $\Phi_{\rm ZPF}^{(m)}$ is the quantum zero-point fluctuations in flux for mode $m$ of the quantum circuit. Using Eqs.~(\ref{cqed1})-(\ref{cqed3}), the interaction Hamiltonian can thus be expressed as
\bea
\hat H_{\rm int} &=& \frac{1}{2 L_{\rm c}} \mleft[ 2 \sum_{n, m} k_n \mleft(\hat c_n +\hat c^\dag_n \mright) \Phi_{\rm ZPF}^{(m)} \mleft( \hat a_m +\hat a_m^\dag \mright) \mright. \nn \\
&&\qquad+ \sum_{m, k} \Phi_{\rm ZPF}^{(m)} \Phi_{\rm ZPF}^{(k)} \mleft( \hat a_m + \hat a_m^\dag \mright) \mleft( \hat a_k + \hat a_k^\dag \mright) \nn\\
&&\qquad+ \mleft. \sum_{n, j} k_n k_j \mleft( \hat c_n + \hat c^\dag_n \mright) \mleft( \hat c_j + \hat c^\dag_j \mright) \mright].
\label{cqed4}
\eea

Neglecting the quadratic terms in the last two lines of \eqref{cqed4} if they can be considered small, or including them in the bare Hamiltonians of the quantum circuits and the photoabsorber, we obtain
\bea
\hat H_{\rm int} &=& \frac{1}{L_{\rm c}} \sum_{n,m} k_n \mleft(\hat c_n + \hat c^\dag_n \mright) \Phi_{\rm ZPF}^{(m)} \mleft( \hat a_m + \hat a_m^\dag \mright) \nn\\
&=& \sum_n g_n \mleft(\hat c_n + \hat c^\dag_n \mright) \hat O,
\label{cqed5}
\eea
where $g_n = k_n / L_{\rm c}$ and $\hat O = \sum_m \Phi_{\rm ZPF}^{(m)} \mleft( \hat a_m + \hat a_m^\dag \mright)$. 

Observing that the operatorial form of the interaction Hamiltonian in \eqref{cqed5} is the same as that given in \eqref{eq:HintSystemAbsorber}, the results from \secref{sec:EP} imply that the probability to absorb a photon from the quantum circuit in the initial state $\ket{E_i}$ is proportional to the mean value of the operator ${\hat x^- \hat x^+}$, where, in this case,
\be
\hat x^+ = \sum_j \sum_{k < j} \sum_m \Phi_{\rm ZPF}^{(m)} \brakket{E_j}{\mleft( \hat a_m + \hat a_m^\dag \mright)}{E_k} \ketbra{E_k}{E_j}.
\label{cqudx+}
\ee
Of course, to find the eigenstates $\ket{E_j}$ of the quantum circuit, a more detailed description of that system is needed. In general, these eigenstates will include contributions from both artificial atoms and resonator modes in the circuit. Thus, the operators $x^+$ and $x^-$ may not correspond to the bare creation and annihilation operators $a_m$ and $a_m^\dag$.

%%%%%%%%%%%%%%%%%%%%%%%%%%%%%%%%%%%%%%%%%%%%%%%%%%%%%%%%%%%%

\section{Analysis of the nature of photons dressing the ground state of the quantum Rabi Hamiltonian}
\label{sec:Rabi}

As another application of our results from \secref{sec:EP}, we now study in more detail the ground state of an ultrastrongly coupled light-matter system. We wish to clarify the question of the virtual nature of excitations in parts of the system that contribute to the ground state of the system as a whole. Using the formalism from \secref{sec:EP}, we will perform a gedanken experiment which, in principle, lets us estimate the lifetime of such excitations.

\subsection{The quantum Rabi model}

We consider the quantum version of the Rabi model~\cite{Rabi1937}, which describes a two-level atom interacting with a single electromagnetic mode. The full system Hamiltonian is
\be
\hat H_{\rm R} = \omega_0 \hat a^\dag \hat a + \frac{\omega_{\rm a}}{2} \sz + g \sx \mleft( \hat a + \hat a^\dag \mright),
\ee
where $a$ ($a^\dag$) is the annihilation (creation) operator for the electromagnetic mode, $\omega_0$ is the resonance frequency of said mode, $\omega_{\rm a}$ is the transition frequency of the two-level atom, $\sz$ and $\sx$ are Pauli matrices, and $g$ is the strength of the light-matter coupling.

If the coupling strength $g$ is much smaller than the resonance frequencies $\omega_0$ and $\omega_{\rm a}$, the RWA can be applied to reduce $\hat H_{\rm R}$ to the Jaynes--Cummings (JC) Hamiltonian~\cite{Jaynes1963}
\be
\hat H_{\rm JC} = \omega_0 \hat a^\dag \hat a + \frac{\omega_{\rm a}}{2} \sz + g \mleft(\sm \hat a^\dag + \sp \hat a \mright),
\ee
where $\sm$ ($\sp$) is the lowering (raising) operator of the atom. The JC Hamiltonian is easy to diagonalize and has the ground state $\ket{g,0}$, where $\ket{g}$ is the ground state of the atom and the second number in the ket indicates the number of photons in the electromagnetic mode. 

However, if the coupling strength increases, the full quantum Rabi Hamiltonian must be used. This Hamiltonian can also be solved~\cite{Braak2011}; the eigenstates can be written in the form
\be
\ket{E_j} = \sum_{k = 0}^\infty \mleft( c^j_{g, k} \ket{g, k} + d^j_{e, k} \ket{e, k} \mright),
\label{rabistate}
\ee
where $\ket{e}$ denotes the excited state of the atom. In particular, the ground state of $\hat H_{\rm R}$ is
\be
\ket{E_0} = \sum_{k = 0}^\infty \mleft( c^0_{g, 2k} \ket{g, 2k} + d^0_{e, 2k+1} \ket{e, 2k+1} \mright)
\label{zerotilde0}
\ee
with non-zero coefficients $c^0_{g, k}$ and $d^0_{e, k}$ for states that contain an even number of bare atomic and photonic excitations. Thus, if we calculate the expectation value of the bare photon number, the result is
\be
\brakket{E_0}{\hat a^\dag \hat a}{E_0} = \sum_{k = 0}^\infty \mleft( 2k \abssq{c^0_{g, 2k}} + \mleft( 2k + 1 \mright)\abssq{d^0_{e, 2k+1}} \mright) \neq 0.
\ee
As mentioned in the introduction, several theoretical studies~\cite{Ciuti2006, Ridolfo2012, Stassi2013} have shown that these photons that are present in the ground state cannot be observed outside the system, since they do not correspond to output photons that can be detected. The diagrammatic approach to the quantum Rabi model in Ref.~\cite{DiStefano2017} also suggests that these photons should be thought of as virtual. However, there are theoretical proposals~\cite{Garziano2014,Lolli2015, Cirio2017} for indirect, non-demolition detection of the photons in the ground state. The question may thus arise whether these photons should be termed virtual or real.

\subsection{Attempting to detect ground-state photons through absorption}

Our approach to photon detection allows us to elucidate the nature of the ground-state photons in the quantum Rabi model in two ways. First, we consider whether the photons can be detected with a photo-absorber. From the treatment in \secref{sec:EP}, we know that a photo-absorber coupled to the electromagnetic mode will have an excitation probability proportional to $\brakket{E_0}{\hat x^- \hat x^+}{E_0}$ when the system described by the quantum Rabi model is in its ground state. Since
\bea
\hat x^+ \ket{E_0} &=& \sum_j \sum_{k < j} \brakket{E_j}{\hat a + \hat a^\dag}{E_k} \ket{E_k}\braket{E_j}{E_0} \nn\\
&=& \sum_{k < 0} \brakket{E_0}{\hat a + \hat a^\dag}{E_k} \ket{E_k} = 0,
\eea
because there are no terms with $k<0$, we conclude that a photo-absorber is not able to detect any photons in $\ket{E_0}$. Note that this is a more general result than what has been obtained with input-output theory. It does not only hold for photon detectors placed outside the resonator hosting the electromagnetic mode; it also holds for photon detectors placed \textit{inside} the interacting light-matter system.

\subsection{Probability of photoabsorption for short times}

The above would seem to further strengthen the case for calling the ground-state photons virtual, but another objection to that would be that virtual particles only exist for very short times, while the excitations considered here are always present in $\ket{E_0}$. As a further application of our approach to photon detection, we therefore calculate the lifetime of the excitations present in the ground state. We begin by noting that the photon-detection theory used here is based on Fermi's golden rule, and as such it gives the probability of photoabsorption for long times. We now extend this theory to short times.

Applying standard first-order perturbation theory and using \eqref{eq:HintSystemAbsorber}, we can calculate the probability that a photon disappears from the state $\ket{E_0}$ and one of the absorber modes is excited. The matrix element describing this process is
\be
W^{(j)}_{k, 0} = g_n \brakket{E_k, 1_n}{\hat c^\dag_n \hat x^+}{E_0, 0},
\ee
where the second label in the kets indicates the absorber state. Due to the presence in $\ket{E_0}$ of states with a non-zero number of photons, this transition matrix element is non-zero. It is interesting to observe that this matrix element would be zero for the JC model, i.e., replacing $\ket{E_0}$ with $\ket{g,0}$. The resulting transition probability in the case of the quantum Rabi Hamiltonian is~\cite{Cohen-Tannoudji1977}
\be
P = \sum_{n, k} \frac{\abssq{W^{(n)}_{k, 0}}}{\hbar} F^2 \mleft( t, \omega^{(n)}_{k, 0} \mright),
\label{probab}
\ee
where $\omega^{(n)}_{k, 0} = \omega_n + \mleft( E_k - E_0 \mright) / \hbar$, and
\be
F (t, \omega) = \frac{\sin (\omega t / 2)}{\omega / 2}.
\ee
If $t$ is sufficiently large, the function $F (t, \omega)$ can be approximated to within a constant factor by the Dirac delta function $\delta (\omega)$. In that case, we obtain that the transition rate for times $t \gg \mleft( E_1 - E_0 \mright)^{-1}$
\be
\frac{\rd P(t)}{\rd t} = \sum_{n, k} \frac{2 \pi}{\hbar} \abssq{W^{(n)}_{k, 0}} \delta \mleft( \omega^{(n)}_{k, 0} \mright).
\ee
Since $\hbar \omega^{(n)}_{k, 0} > \mleft( E_1 - E_0 \mright)$ is strictly larger than zero and is of the order of $\omega_0$, no transitions will be observed for large times. However, for $t \ll 1/ \omega^{(n)}_{k, 0}$,
\be
P = \sum_{n, k}	\frac{\abssq{W^{(n)}_{k, 0}}}{\hbar} t^2,
\ee
and thus the photons in $\ket{E_0}$ can induce transitions with a very small probability (due to the $t^2$ term) during a small time interval. This means that the ground-state photons are coming into existence for very short times, on the order of a period of the electromagnetic mode, in agreement with the time-energy uncertainty principle. This is consistent with the interpretation of the ground-state photons as virtual rather than real.

%%%%%%%%%%%%%%%%%%%%%%%%%%%%%%%%%%%%%%%%%%%%%%%%%%%%%%%%%%%%
   
\section{Conclusions}
\label{sec:Conclusions}

We have explored photon detection for quantum systems with arbitrarily strong light-matter interaction. In these systems, the very strong interaction makes light and matter hybridize such that a naive application of standard photodetection theory can lead to unphysical results, e.g., photons being emitted from the ground state of a system. While some previous works have shown how to amend input-output theory to arrive at correct expressions for the photon output flux, we have presented a more complete theory for photon detection in these systems without relying on input-output theory. We followed Glauber's original approach for describing photon detection and found, using Fermi's golden rule, the correct excitation probability rate for a photoabsorber interacting with the light-matter quantum system. Calculating this rate requires knowledge of the system eigenstates, such that the system operator coupling to the photoabsorber can be divided into negative- and positive-frequency components. The difference with standard photon detection arises because the strong light-matter interaction dresses the system states such that the aforementioned components no longer correspond to bare annihilation and creation operators.

We presented results for both wide-band and narrow-band photon detectors. We then showed in detail how the formalism can be applied to two representative quantum systems that can display strong light-matter interaction: cavity QED with an atom acting as the photoabsorber, and a circuit-QED setup with inductive coupling to a photon detector. Although the results we derived here were limited to second-order correlation functions, they can be directly generalized to higher-order normal-order correlation functions.

We also applied our photon-detection formalism to the quantum Rabi Hamiltonian, which describes a two-level atom interacting with a single electromagnetic mode. For large light-matter interaction, the ground state of this model contains photons, and whether these photons are virtual or real has been subject to debate. Using our formalism, we were able to clarify the nature of the ground-state photons in two ways. First, we showed that the ground-state photons, in the limit of long times, will not be detected by a photoabsorber, no matter where this photoabsorber is placed. Unlike previous results obtained with modified input-output theory, our result also holds for a detector placed inside the system (e.g., inside an optical cavity). Second, we considered a gedanken experiment, where the excitation probability of the photoabsorber is calculated for short times. We found that there is a small excitation probability for such short times, which is consistent with an interpretation where virtual photons are flitting in and out of existence on a time-scale set by the time-energy uncertainty relation. Our results thus provide further evidence for the virtual nature of the photons present in the ground state of the quantum Rabi Hamiltonian.  

\bibliography{References_PhotodetectionUSC}

\end{document}